\documentstyle[12pt]{article}
\begin{document}
\title{Geometry and Quantum Mechanics}
\author{B.G. Sidharth\\
B.M. Birla Science Centre, Adarshnagar, Hyderabad - 500 063,
India}
\date{}
\maketitle
\begin{abstract}
Attempts for a geometrical interpretation of Quantum Theory were
made, notably the deBroglie-Bohm formulation. This was further
refined by Santamato who invoked Weyl's geometry. However these
attempts left a number of unanswered questions. In the present
paper we return to these two formulations, in the context of
recent studies invoking fuzzy spacetime and noncommutative
geometry. We argue that it is now possible to explain the earlier
shortcomings. At the same time we get an insight into the
geometric origin of the deBroglie wavelength itself as also the
Wilson-Sommerfeld Quantization rule.
\end{abstract}
\section{Introduction}
One of the fruitful approaches to Quantum Mechanics was the so
called deBroglie-Bohm hydrodynamical formulation \cite{r1}, which
originated with Madelung and was developed by Bohm using
deBroglie's pilot wave ideas. In this formulation, while the
initial position coordinates in a Quantum Mechanical trajectory,
are random, the trajectories themselves are determined by
classical mechanics. Quantum Mechanics enters through an
inexplicable Quantum potential which is again related to the wave
function. This has been a stumbling
block in the acceptance of the formulation.\\
Much later, Santamato further developed the deBroglie-Bohm
formulation by relating the mysterious Quantum potential to
fundamental geometric properties, by invoking Weyl's geometry
\cite{r2,r3,r4}. The net result was that the mysterious Quantum
effects were shown to be related to the geometric structure of
space specifically to the curvature. Unfortunately, Weyl's theory
itself did not find favour \cite{r5}. Apart from anything else,
the theory sought to unify electromagnetism with gravitation, but
on closer scrutiny, in this geometrical structure the two
interactions were actually independent and ad hoc entities as
noted
by Einstein himself \cite{r6}.\\
We will now reexamine all this in the light of recent developments
in fuzzy spacetime and noncommutative geometry, and argue that
infact, once the underlying fuzzyness is recognized, then the
above apparent difficulties disappear.
\section{The deBroglie-Bohm Formulation and Extensions}
Let us briefly review the above theory \cite{r7}. We start with
the Schrodinger equation
\begin{equation}
\imath \hbar \frac{\partial \psi}{\partial t} = -
\frac{\hbar^2}{2m} \nabla^2 \psi + V \psi\label{e1}
\end{equation}
In (\ref{e1}), the substitution
\begin{equation}
\psi = Re^{\imath S/\hbar}\label{e2}
\end{equation}
where $R$ and $S$ are real functions of $\vec r$ and $t$, leads
to,
\begin{equation}
\frac{\partial \rho}{\partial t} + \vec \nabla \cdot (\rho \vec v
) = 0\label{e3}
\end{equation}
\begin{equation}
\frac{1}{\hbar} \frac{\partial S}{\partial t} + \frac{1}{2m} (\vec
\nabla S)^2 + \frac{V}{\hbar^2} - \frac{1}{2m} \frac{\nabla^2
R}{R} = 0\label{e4}
\end{equation}
where
$$\rho = R^2 , \vec v = \frac{\hbar}{m} \vec \nabla S$$
and
\begin{equation}
Q \equiv - \frac{\hbar^2}{2m} (\nabla^2 R/R)\label{e5}
\end{equation}
Using the theory of fluid flow, it is well known that (\ref{e3})
and (\ref{e4}) lead to the Bohm alternative formulation of Quantum
Mechanics. In this theory there is a hidden variable namely the
definite value of position while the so called Bohm potential $Q$
can be non local, two features which
do not find favour with physicists.\\
Let us now briefly review Weyl's ideas. He postulated that in
addition to the general coordinate transformations of General
Relativity, there were also gauge transformations which multiplied
all components of the metric tensor $g_{\mu \nu}$, by an arbitrary
function of the coordinates. So, the line elements would no longer
be invariant. In its modern version, the metric tensor is
normalized so that its determinant is given by \cite{r5},
$$|g_{\mu \nu} | = -1,$$
while it now transforms as a tensor density of weight minus half,
and not as a tensor. This leads to the circumstance that there is
now a covariant derivative involving an arbitrary function of
coordinates $\Phi_\mu$ given by
\begin{equation}
\Phi_\sigma = \Gamma^\rho_{\rho \sigma},\label{e6}
\end{equation}
where the $\Gamma$'s denote the Christofell symbols. Weyl
identified $\Phi_\mu$ in (\ref{e6}) with the electromagnetic
potential. It must be noted that in Weyl's geometry, even in a
Euclidean space there is a
covariant derivative and a non vanishing curvature $R$.\\
Santamato exploits this latter fact, within the context of the
deBroglie-Bohm theory and postulates a Lagrangian given by
$$L (q,{\dot q} , t) = L_c(q, {\dot q} ,t) + \gamma
(\hbar^2/m)R(q,t),$$ He then goes on to obtain the equations of
motion like (\ref{e1}),(\ref{e2}), etc. by invoking an Averaged
Least Action Principle
$$I (t_0,t_1) = E \left\{ \int^t_{t_0} L^* (q(t,\omega),{\dot q} (t,\omega),t)dt\right\}$$
\begin{equation}
= \mbox{minimum} ,\label{e7}
\end{equation}
with respect to the class of all Weyl geometries of space with
fixed metric tensor. This now leads to the Hamilton-Jacobi
equation
\begin{equation}
\partial_t S + H_c (q,\nabla S,t) - \gamma (\hbar^2 /m) R = 0,\label{e8}
\end{equation}
and thence to the Schrodinger equation (in curvi-linear
coordinates)
$$\imath \hbar \partial_t \psi = (1/2m)\left\{ [(\imath \hbar /\sqrt{g})
\partial_\imath \sqrt{g} A_\imath ]g^{\imath k} (\imath \hbar \partial_k
+ A_k)\right\} \psi$$
\begin{equation}
+ [V - \gamma (\hbar^2 /m){\dot R} ] \psi = 0,\label{e9}
\end{equation}
As can be seen from the above, the Quantum potential $Q$ is now
given in
terms of the scalar curvature $R$.\\
We would now like to relate the arbitrary functions $\Phi$ of
Weyl's
formulation with a noncommutative spacetime geometry, as was shown recently.\\
Let us write the product $dx^\mu dx^\nu$ of the line element,
$$ds^2 = g_{\mu \nu} d dx^\mu dx^\nu ,$$
as a sum of half its anti-symmetric part and half the symmetric
part. The line element now becomes $(h_{\mu \nu} + \bar h_{\mu
\nu}) dx^\mu dx^\nu$. This leads us back to the Weyl geometry
because the metric tensor $\bar h_{\mu \nu}$
now becomes a tensor density \cite{r8,r9}.\\
In other words it is the underlying fuzzyness of space time as
expressed by
\begin{equation}
[dx^\mu , dx^\nu ] \approx l^2 \ne 0\label{e10}
\end{equation}
$l$ being a typical length scale $\sim 0(dx^\mu)$, that brings out
Weyl's geometry, not as an ad hoc feature, but as a truly
geometrical consequence, and therefore also legitimises
Santamato's postulative
approach of extending the deBroglie-Bohm formulation.\\
At an even more fundamental level, this formalism gives us the
rationale for the deBroglie wave length itself. Because of the
noncommutative geometry in (\ref{e10}) space becomes multiply
connected, in the sense that a closed circuit cannot be shrunk to
a point within the interval. Let us consider the simplest case of
double connectivity. In this case, if the interval is of length
$\lambda$, we will have, using (\ref{e5}),
\begin{equation}
\Gamma \equiv \int_c m\vec V \cdot d \vec r = h \int_c \vec \nabla
S \cdot d \vec r = h \oint dS = mV \pi \lambda = \pi h\label{e11}
\end{equation}
whence
\begin{equation}
\lambda = \frac{h}{mV}\label{e12}
\end{equation}
In (\ref{e11}), the circuit integral was over a circle of diameter
$\lambda$. Equation (\ref{e12}) shows the emergence of the
deBroglie wavelength. This follows from the noncommutative
geometry of space time, rather than the physical Heisenberg
Uncertainty Principle. Remembering that $\Gamma$ in (\ref{e11})
stands for the angular momentum, this is also the origin of the
Wilson-Sommerfeld quantization rule, an otherwise mysterious
Quantum Mechanical prescription.
\section{Discussion}
1. We would like to stress that Santamato's treatment via Weyl's
geometry, of the deBroglie-Bohm formulation was postulative (Cf.
equations (\ref{e7}), (\ref{e8}), (\ref{e9})), while the Weyl
formulation itself had not found favour for its original
motivation. Perhaps this was the reason why Santamato's
formulation was not taken so seriously. On the other hand, we have
argued from the point of view of the noncommutative geometry
(\ref{e10}), which, after many decades, is now coming to be
recognized in the context of Quantum Superstring theory and
Quantum Gravity.\\
2. It is well known that the so called Nelsonian stochastic
process resembles the deBroglie-Bohm formulation, with very
similar equations \cite{r10,r11}. However in this former case,
both the position and velocity are not deterministic because of an
underlying Brownian process. In this formulation the diffusion
constant of the theory has to be identified with,
$$\nu = \frac{h}{m}$$
These are the extra features in this stochastic formulation,
rather than the Quantum potential, which also appears in the
equations. It has been shown by the author \cite{r12,r7}, that
both the similar approaches infact can be unified for relativistic
velocities, by considering quantized vortices originating from
(\ref{e11})of the order of the deBroglie, now the Compton scale
$l$. This immediately brings us back to the fuzzy noncommutative
geometry (\ref{e10}). At the same time it must be pointed out that
the supposedly unsatisfactory non local features of the Quantum
potential $Q$ become meaningful in the above context at the
Compton scale, within which indeed we
have exactly such non local effects \cite{r13}.\\
It may be pointed out that more recently we have been led back to
the background Quantum vaccuum and the underlying Zero Point
Field, now christened dark energy by the observation of the
cosmological constant implied by the accelerated expansion of the
universe \cite{r7} and it is this ZPF which provides the Brownian
process of the stochastic theory. As pointed out by Nottale
\cite{r14}, such a Brownian process also eliminates the hidden
variable feature of the deBroglie-
Bohm formulation.\\
Interestingly it has been argued by Enz \cite{r15} that a particle
extension, as is implied in the above considerations in the form
of a quantized vortex or fuzzy space time, is the bridge between
the particle and wave aspects. Infact at this scale, there is
zitterbewegung, reminiscent of deBroglie's picture of a particle
as an oscillator. Originally Dirac had interpreted the
zitterbewegung oscillations as unphysical, which are removed by
the fact that due to the Uncertainty Principle we cannot go down
to arbitrarily small space time intervals, so that our space time
points are averages over Compton scale intervals (Cf.\cite{r7,r16}
for a discussion).\\
3. It is also interesting to note that Santamato's tying up of the
Quantum potential with the curvature $R$ has been interpreted as
being the result of cosmic fluctuations \cite{r17}, this being a
special case of a more general but identical earlier result of the
author \cite{r18,r19}.\\
4. Finally we observe that a complete geometrification of Quantum
Theory is now possible. Via equation (\ref{e2}) we have introduced
in the wave function the (mass) density $\rho$ and the phase $S$
of the wave function $\psi$. As we have seen the phase $S$ is
directly related to the determinant of $|g|$ infact we have
$$\vec \nabla S = \vec \nabla (ln\sqrt{|g|}$$
Further in the above formulation (Cf.ref.\cite{r2}) infact we have
$$\Phi_\imath = \vec \nabla (ln \rho )$$
Thus the mass and gravitational and electromagnetic forces (as
also the strong and electroweak interactions Cf.ref.\cite{r7}),
are related to the basic metrical properties of spacetime.

\end{document}